\documentclass[aps,pra,letterpaper,10pt,twocolumn]{revtex4-1}
\usepackage[colorlinks=true,allcolors=blue]{hyperref}
\usepackage{amssymb}
\usepackage{amsmath}
\usepackage{graphicx}
\usepackage[caption=false]{subfig}
\usepackage{amsfonts}
\usepackage{xcolor}
\usepackage{epsfig}
\usepackage{color}
\usepackage{bm}
\usepackage{tabularx}
\usepackage{multirow}
\usepackage{mathtools}
\usepackage{xfrac}
\usepackage{blkarray}
\usepackage{bbold}
\usepackage[mathscr]{euscript}
\usepackage{autobreak}
\usepackage{comment}
\usepackage[LY1]{fontenc}

\newcommand{\vn}{V$_{\rm N}$}
\newcommand{\vb}{V$_{\rm B}$}

\newcommand{\bn}{B$_{\rm N}$}
\newcommand{\nb}{N$_{\rm B}$}
\newcommand{\bnnb}{B$_{\rm N}$N$_{\rm B}$}

\begin{document}
\title{Native antisite defects in $h$-BN}

\author{Song Li}
\email[Electronic mail: ]{li.song@csrc.ac.cn}
\affiliation{Beijing Computational Science Research Center, Beijing 100193, China}
\affiliation{HUN-REN Wigner Research Centre for Physics, Institute for Solid State Physics and Optics, P.O.\ Box 49, H-1525 Budapest, Hungary}

\author{Pei Li}
\affiliation{School of Integrated Circuit Science and Engineering, Tianjin University of Technology, Tianjin 300384, China}
\affiliation{Beijing Computational Science Research Center, Beijing 100193, China}

\author{Adam Gali}
\email[Author to whom correspondence should be addressed: ]{adam.gali@wigner.hun-ren.hu}
\affiliation{HUN-REN Wigner Research Centre for Physics, Institute for Solid State Physics and Optics, P.O.\ Box 49, H-1525 Budapest, Hungary}
\affiliation{Department of Atomic Physics, Institute of Physics, Budapest University of Technology and Economics, M\H{u}egyetem rakpart 3., H-1111 Budapest, Hungary}
\affiliation{MTA-WFK "Lend\"ulet" Momentum Semiconductor Nanostructures Research Group, P.O.\ Box 49, H-1525 Budapest, Hungary}

\date{\today}
\begin{abstract}
Hexagonal boron nitride (hBN) is an excellent host for solid-state single phonon emitters. Experimental observed emission ranges from infrared to ultraviolet. The emission centers are generally attributed to either intrinsic or extrinsic point defects embedded into hBN. Nevertheless, the microscopic structure of most of these defect emitters is uncertain. Here, through density-functional theory calculations we studied the native antisite defects in hBN. We find that the neutral boron antisite might be a nonmagnetic single photon source with zero-phonon-line (ZPL) at 1.58~eV and such a lineshape that is often observed in experiments. Furthermore, the positively charged nitrogen antisite might be associated with a dim color center recently observed as a blue emitter with ZPL at 2.63~eV. These simple single substitution defects indicate the existence of out-of-plane phonon mode which significantly affects the optical properties. Our results could provide useful information for identification of quantum emitters in hBN.   
\end{abstract}

\maketitle

%
%

The wide band gap of hexagonal boron nitride (hBN) is a promising host material for bound states induced by optically active deep defect levels acting as isolated two-level system ~\cite{wolfowicz2021quantum, zhang2020material} for quantum applications including nanoscale sensing, computing and information processing~\cite{tran2016quantum, gottscholl2020initialization, chejanovsky2021single, mendelson2021identifying, hayee2020revealing, bourrellier2016bright, bommer2019new, tran2016robust}. The fabrication and isolation of these defects is crucial for generating single photon emitters for quantum information processing applications. Numerous experiments have reported defects in hBN responsible for quantum emission with wavelength across visible and ultraviolet range~\cite{tran2016robust,vuong2016phonon,bourrellier2016bright,tran2016quantum, gottscholl2020initialization, chejanovsky2021single, mendelson2021identifying}. Surprisingly, many emitters in hBN demonstrate high brightness and robustness at room temperature with narrow linewidth.

Deterministic creation of defects in hBN has been achieved in atomic precision which is a prerequisite for scalable integration of emitters with photonic circuits and architectures~\cite{xu2021creating, gale2022site, ziegler2019deterministic}. Recent high resolution transmission electron microscopy (HRTEM) images could demonstrate the defect morphology on the very top layer of hBN~\cite{jin2009fabrication, krivanek2010atom, hayee2020revealing}. However, no direct correlation of defect location and emission from confocal microscope photoluminescence (PL) measurements is expected as the typical spatial resolution of the confocal microscope is limited to hundreds nanometers, therefore quantum emitters cannot be distinguished from deeper regions and the top layer of multilayer hBN. 

\begin{figure}[tb]
\includegraphics[width=0.9\columnwidth]{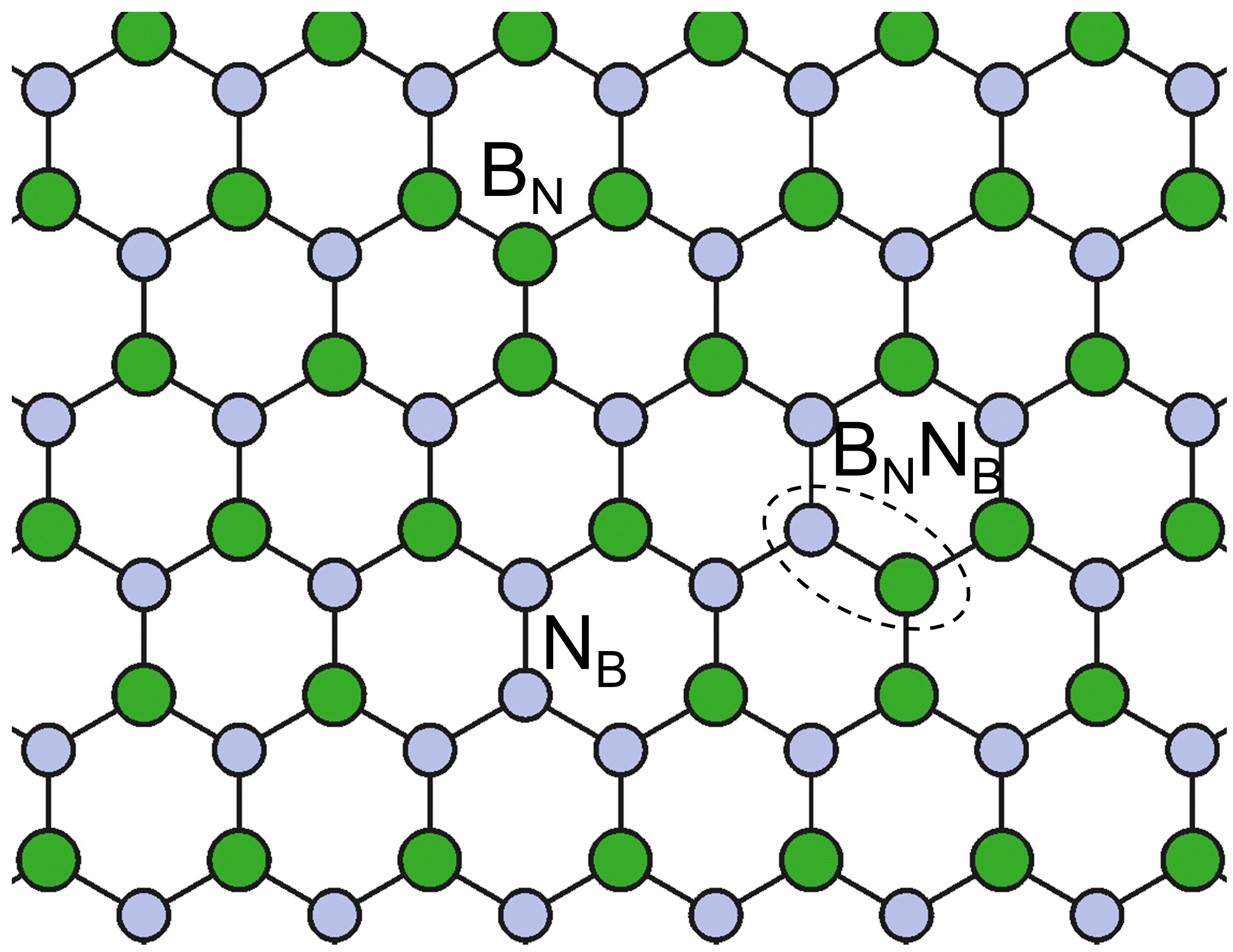}
\caption{\label{Figure1}%
Schematic view of native antisite defects \bn, \nb, and \bnnb.}
\end{figure}

Series of studies proposed numerous defect models and assigned them to defect emitters, including native vacancy defects like boron-vacancy (\vb) ~\cite{abdi2018color, ivady2020ab, reimers2020photoluminescence}, nitrogen-vacancy (\vn)~\cite{sajid2018defect, su2022tuning}, and Stone-Wales (SW) defects~\cite{tawfik2017first, hamdi2020stone}; external impurities, especially carbon defects~\cite{sajid2018defect, weston2018native, chejanovsky2021single, mendelson2021identifying, auburger2021towards, mackoit2019carbon, li2022bistable} and oxygen-related defects~\cite{tan2022donor,li2022identification,li2024quantum}. The robustness of the emission lines against various fabrication methods of hBN indicates the responsible defects are highly stable and native. Although the native vacancies are investigated before, we find the native antisite defects are seldom considered, as shown in Fig.~\ref{Figure1}.

In this Letter, we systematically study the magneto-optical properties of native antisite defects in hBN. We determine the PL spectrum and optical lifetime, and find good agreement with the properties of previously observed quantum emitters. Our results here indicate that the antisite defects might exist and act as single photon sources in hBN. 

\begin{figure}[tb]
\includegraphics[width=\columnwidth]{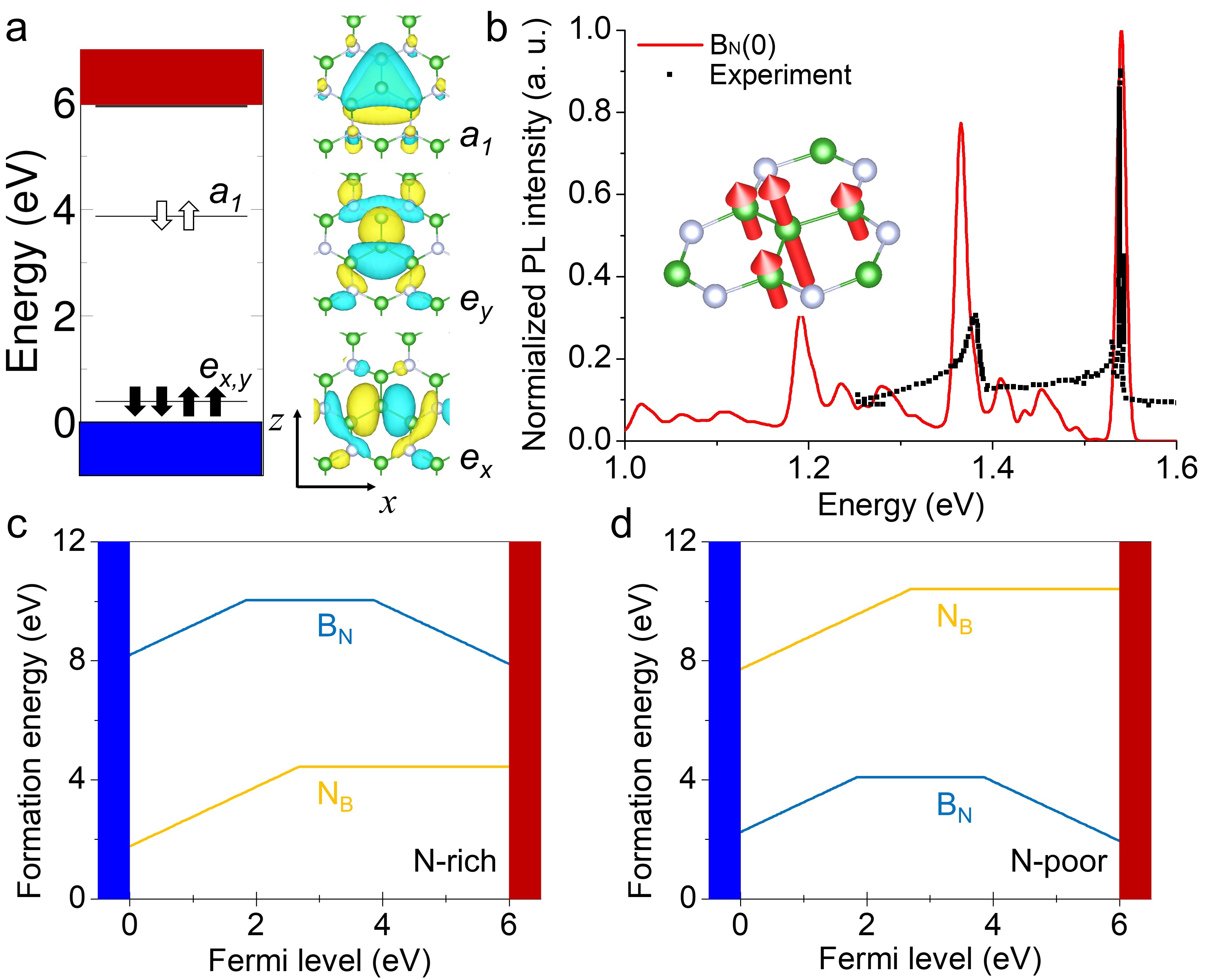}
\caption{\label{Figure2}%
(a) Ground state energy level of \bn($0$) defect with $C_{3v}$ symmetry in hBN. (b) The experimental (Ref.~\citenum{huang2022carbon}) and simulated PL spectrum. The inset shows the vibration associated with the JT distortion. (c,d) The formation energy of \bn\ and \nb\ under N-rich and poor condition.}
\end{figure}

With the plane-wave basis \textit{Vienna ab initio simulation package} (VASP) code~\cite{kresse1996efficiency, kresse1996efficient}, we carry out spin-resolved DFT calculations for the antisite defects in hBN. The valence electrons are separated from core nuclei with projector augmented wave (PAW) formalism~\cite{blochl1994projector, kresse1999ultrasoft}. A $8\times8$ two-layered bulk model with 256 atoms is used to avoids artificial defect-defect interaction. We use DFT-D3 method of Grimme~\cite{grimme2010consistent} for dispersion correction due to interlayer vdW interaction. A modified screened hybrid density functional of Heyd, Scuseria, and Ernzerhof (HSE)~\cite{heyd2003hybrid} is used to optimize the structure and calculate the electronic properties with mixing parameter $\alpha = 0.32$ for the Hartree–Fock exchange. The structure is fully relaxed until the force on atoms are less than 0.01~eV/\AA\ and the cutoff energy is 450~eV. The single $\Gamma$-point scheme is applied for supercell calculation. $\Delta$SCF method~\cite{gali2009theory} is used to calculate excited states. The PL spectrum is simulated based on Franck-Condon approximation at 0~K which simplifies the simulation of the PL lineshape to the overlap between the phonon modes in the electronic ground and excited states~\cite{gali2009theory,alkauskas2012first}. 

The defect formation energy $E_\text{f}$ is calculated to determine the charge stability as follows,
\begin{equation}
\begin{split}
E^q_\text{f} = &E^q_\text{d} - E_\text{per} -\sum_i n_i\mu_i + q\left(E_\text{VBM} + E_\text{Fermi}\right)\\
&+ E_\text{corr}\left(q\right)\text{,}
\end{split}
\end{equation}
where $E_\text{d}^q$ is the total energy of hBN model with defect at $q$ charge state and $E_\text{per}$ is the total energy of hBN layer without defect. $n_i$ denotes how many atoms that are added ($n>$ 0) or removed ($n<$ 0) from perfect supercell, and $\mu_i$ is the chemical potential of boron or nitrogen atom. We use the chemical potential $\mu_N$ from nitrogen molecule for the N-rich condition and chemical potential $\mu_B$ from bulk boron for N-poor condition. In thermodynamic condition, the bounds are set by $\mu_N$ + $\mu_B$ = $\mu_{BN}$. The chemical potentials provide boundary cases while they cannot represent experimental conditions during growth. The Fermi level $E_\text{Fermi}$ represents the chemical potential of electron reservoir and it is aligned to the valence band maximum (VBM) energy of perfect hBN, $E_\text{VBM}$. The $E_\text{corr}\left(q\right)$ is the correction term for the charged supercell due to the existence of electrostatic interactions with periodic condition.
The charge correction is done by SXDEFECTALIGN code~\cite{freysoldt2018first}. We applied this correction to such special excited states of the neutral defects where band edge excitation is involved (see Ref.~\citenum{gali2023recent} for explanation).

The \bn\ is nonmagnetic in the neutral charge state. There are empty $a_1$ states with degenerate $e$ states in gap. We find that the in-plane configuration is a dynamic Jahn-Teller (DJT) system as explained below. Our phonon calculation yields one imaginary phonon around 16~meV and the visualization of this mode is shown in Fig.~\ref{Figure2}b. It corresponds to an out-of-plane local vibration mode in which the antisite boron atom is mostly involved. This means that the \bn\ defect prefers out-of-plane distortion with $C_{3v}$ symmetry rather than the planar $D_{3h}$ symmetry. Considering the horizontal mirror plane, we can imagine there is another equilibrium out-of-plane configuration. The out-of-plane displacement of the central boron atom is 0.63~\AA. The calculated energy difference is 59~meV with PBE functional while it is 2~meV with HSE functional. It exhibits the character of the DJT systems where the central atom of the defect can tunnel between two equivalent configurations with small energy barrier. We note that recent experiments already confirmed the existence of the out-of-plane distortion for certain quantum emitters and it could lead to decoupling from in-plane phonon mode~\cite{hoese2020mechanical}. We then calculate the ZPL energy by promoting an electron from the $e$ state to the $a_1$ state in the gap. With symmetry conserving excitation, the ZPL energy is 1.74~eV with Huang-Rhys (HR) factor $S= 1.09$; however, the high symmetry configuration of excited state is JT unstable due to the half occupation of the $e$ state. The central boron atom is pushed out along B-B bond direction so the degeneracy then would be lifted and the symmetry reduces to $C_{2v}$ to lower the total energy. Hence, the final ZPL energy is 1.58~eV with enhanced $S = 1.68$ due to symmetry breaking, as summarized in Table \ref{tab:data}. The nonmagnetic quantum emitters have been observed in hBN with ZPL energy around 1.6~eV (known as type-II emitters)~\cite{li2017nonmagnetic}. However, type-II emitters exhibit extremely small sideband, indicating very weak electron-phonon coupling in the optical transition. Beside that, another PL spectra is recorded in hBN thin film with ZPL energy around 1.55~eV~\cite{huang2022carbon}. In Fig.~\ref{Figure2}b we display the comparison between our simulated and the experimental PL spectra at 1.6~K. Although the data of low energy range is missing, the location of simulated first phonon replica is consistent with the observed one. Thus, we tentatively associate \bn($0$) with this emitter.

\begin{figure}
\includegraphics[width=\columnwidth]{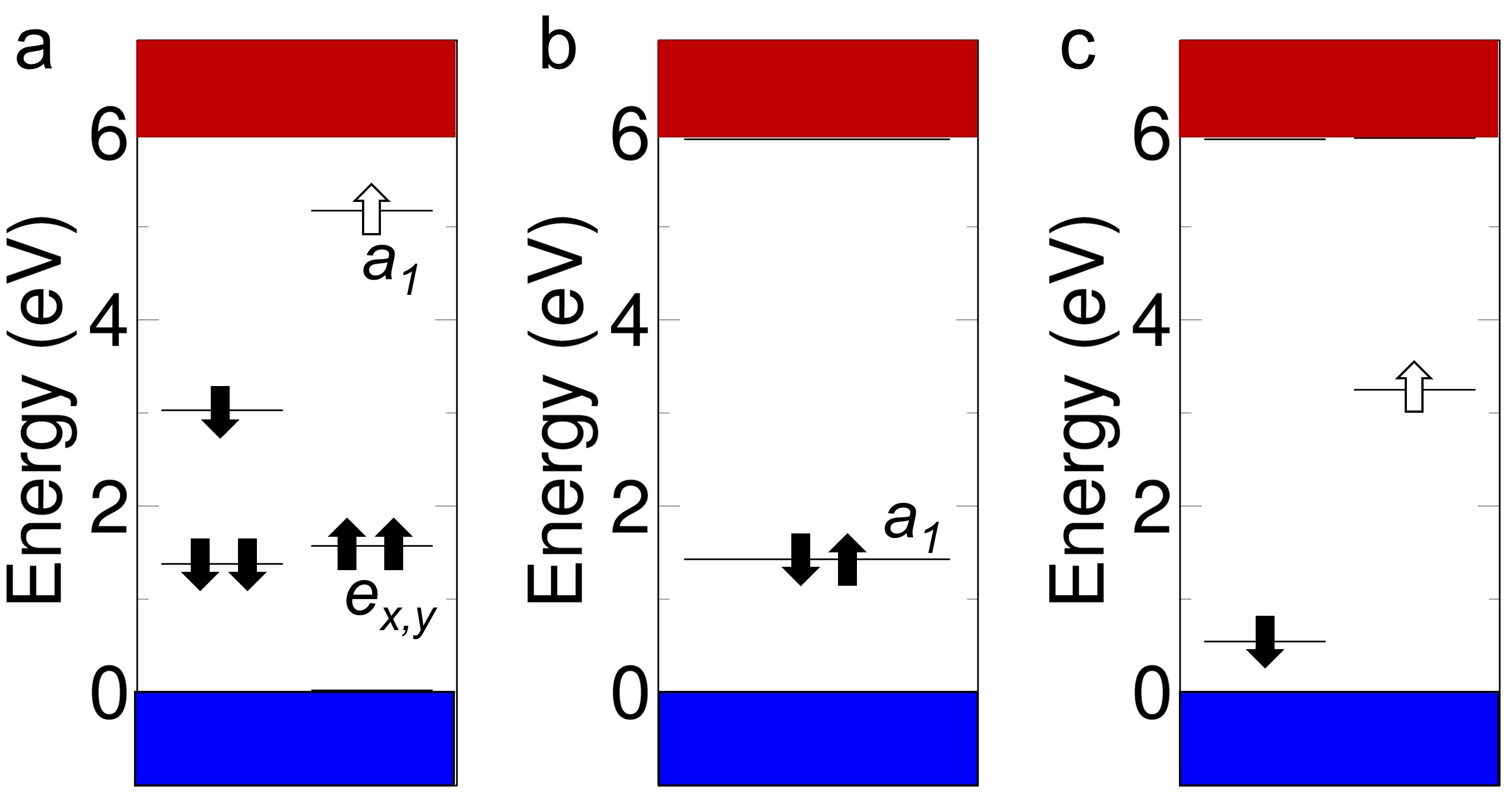}
\caption{\label{Figure3}%
Energy level diagram of (a) \bn($-$) (b) \nb($0$) and (c) \nb($+$) in the ground state. The filled and empty arrows indicate the occupied and unoccupied defect states in the respective spin-up and spin-down channels.}
\end{figure}

Formation energy calculations indicate that \bn\ can be positively or negatively charged with charge transition level (CTL) at $E_\text{VBM}$ $+$ 1.84 and 3.86~eV for ($+1/0$) and ($0/-1$), respectively, and the difference in the results from a previous study~\cite{weston2018native} is due to the geometry distortion considered here. The \bn($-$) also prefers out-of-plane configuration in the ground state. There exist two possible optical transitions as the $a_1$ level is filled with one electron: from $a_1$ to conduction band minimum (CBM) in the spin majority channel and $e$ to $a_1$ in the spin minority channel, as shown in Fig.~\ref{Figure3}a. Our calculation indicate the latter one has 0.49~eV smaller ZPL energy than the first one. The final ZPL energy is 1.75~eV with $S = 2.13$. We predict the \bn($+$) should have similar properties as it has the same type of optical transition but the JT distortion happens in the ground state while retrieve a $C_{3v}$ symmetry in the excited state.

The \nb($0$) is nonmagnetic with a fully occupied $a_1$ state in the gap, as depicted in Fig.~\ref{Figure3}b. The energy difference between in-plane and out-of-plane configurations is 0.55~eV with displacement at around 0.45~\AA. The possible optical transition is from $a_1$ to the conduction band edge. Surprisingly, we find the excitation process annihilate the out-of-plane relaxation and the geometry goes back to high symmetry $D_{3h}$. This is a pseudo JT system discussed before~\cite{li2020giant}. This effect possibly generates a PL center with ZPL energy at around 3.11~eV and a huge electron-phonon coupling with $S = 21.6$.

\nb($+$) is stable when the Fermi-level is below $E_\text{VBM}$ $+$ 2.68~eV (see Ref.~\citenum{weston2018native}). In this charge state, the $D_{3h}$ symmetry is kept similar to the neutral excited state. The lowest optical transition is from VBM to defect level. The calculated ZPL energy is 2.63~eV with $S = 2.13$. Recent experiments report the emission energy around 435~nm (2.8-2.9~eV) which are called blue emitters in hBN~\cite{shevitski2019blue, gale2022site, fournier2021position, zhigulin2022stark}. One blue emitter shows non-linear Stark shift with small transition dipole moment (0.1~Debye)~\cite{zhigulin2022stark}. The absence of a permanent dipole moment is attributed to the high symmetry of the defect with inversion symmetry. Single atom substitution with $D_{3h}$ symmetry is the simplest model of such system. \nb($+$) does not introduce degenerate states in the gap therefore there is no symmetry breaking due to JT instability and it has the feature of no permanent dipole moment. Recent experiment observed PL peak at 2.6~eV and it is robust against annealing~\cite{dkabrowska2024defects}. Based on these facts and the calculated properties of the defect model, we speculate that \nb($+$) is a reasonable candidate for the origin of the dim blue single photon emitter in hBN.

Under N-rich condition, the N-related defects \nb\ and interstitial N$_\text{i}$ have the lowest formation energy. The different charge states of these two defects could pin the Fermi level at the crossing point of formation energy of \nb\ and N$_\text{i}$, which is 2.18 eV above VBM in Ref.~\cite{weston2018native} while could be further shifted to around 3 eV if we consider another possible configuration of N$_\text{i}$~\cite{khorasani2021identification}. In either circumstances, the Fermi level is pinned to the middle of the gap and \bn\ has a formation energy at around 4~eV that results in relatively low concentration of these defects. The position of Fermi level was also analyzed in N-poor condition and the result varies depending on the defects considered to maintain charge neutrality~\cite{maciaszek2022thermodynamics,weston2018native}. The formation energy of \bn\ is similarly high so it is neither an abundant defect when created under quasi thermodynamic equilibrium conditions.
We note that the single photon emitters could be only observed if the concentration of the underlying defects is relatively low because of the relatively low spatial resolution of the traditional optical methods, e.g., confocal microscope. Thus, the high formation energy of \nb($+$) could explain that the dim blue emitter is observed as a single photon emitter and not reported as a common emitter. We apply the same argument for the \bn($0$) model that it can well explain certain near infrared quantum emitters in hBN.

\renewcommand{\arraystretch}{1.5}
\begin{table}
\caption{\label{tab:data} Calculated ZPL energy and HR factor for antisite defects with high or low symmetry cases.}
\begin{ruledtabular}
\begin{tabular}{l|ccc}
Defect & Symmetry & ZPL (eV) & HR factor\\
\hline
\bn($0$) & $D_{3h}$ & 1.71 & 0.42 \\
\bn($0$) & $C_{3v}$ & 1.74 & 1.09 \\
\bn($0$) & $C_{3v}$-JT & 1.58 & 1.68 \\
\bn($-$) & $D_{3h}$ & 1.94 & 0.49 \\
\bn($-$) & $C_{3v}$ & 1.82 & 1.46 \\
\bn($-$) & $C_{3v}$-JT & 1.75 & 2.13 \\
\nb($0$) & $D_{3h}$ & 2.56 & -- \\
\nb($0$) & $C_{3v}$ & 3.11 & 21.6 \\
\nb($+$) & $D_{3h}$ & 2.63 & 2.13 \\
\bnnb($0$) & $C_{2v}$ & 2.98 & 2.58 \\

\end{tabular}
\end{ruledtabular}
\end{table}

\begin{figure}[htbp]
    \includegraphics[width=\columnwidth]{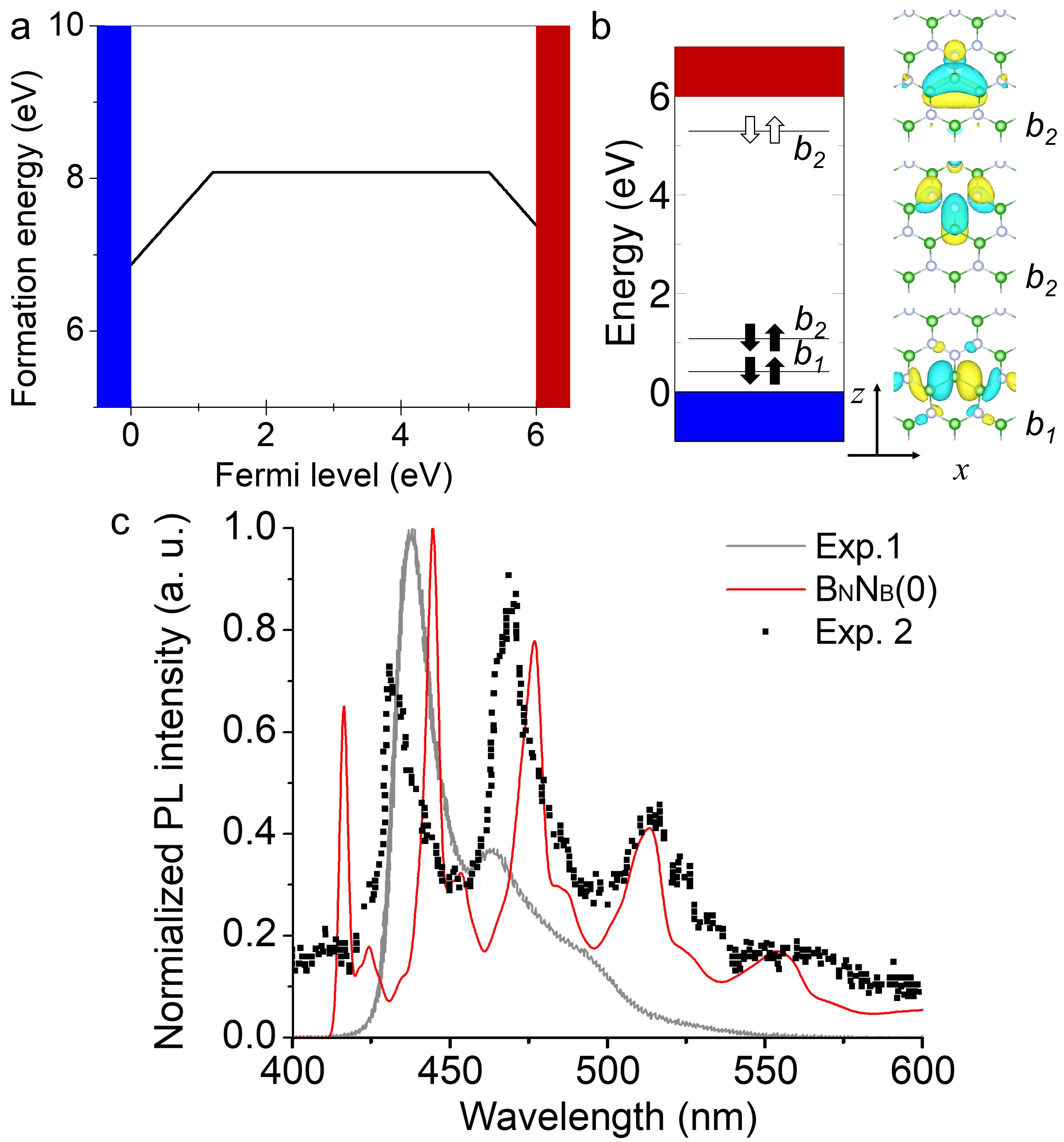}
    \caption{\label{Figure4}(a) Formation energy as a function of Fermi level of \bnnb\ defect. (b) Energy level diagram of \bnnb\ and the wavefunction of defect orbitals. (c) The experimental (Exp.\ 1~ from Ref.~\citenum{gale2022site} and Exp.\ 2 from Ref.~\citenum{liang2023blue}) and simulated PL spectra.}
\end{figure}

We here also considered the antisite pairs as the combination of \bn\ and \nb\, namely \bnnb\ that nitrogen and boron atoms exchange their position. The formation energy of \bnnb\ does not depend on the chemical potential of boron or nitrogen. The formation energy calculation indicates a stable neutral charge state with high formation energy over 8~eV (Fig.~\ref{Figure4}a) so the concentration of such defect is extremely low in thermodynamic equilibrium condition and could be generated by irradiation. The energy level of \bnnb\ is plotted in Fig.~\ref{Figure4}b. It is nonmagnetic in the neutral charge state. With $C_{2v}$ symmetry, we can identify occupied $b_1$ and $b_2$ orbitals and an unoccupied $b_2$ orbital in the gap. Naturally, the lowest possible optical transition is between the two $b_2$ orbitals. This is a bright transition since the orbitals share the same character. The calculated optical transition dipole moment ($\mu$) is 3~Debye, much larger than that of \nb($+$). Based on this result, the radiative lifetime can be evaluated as
\begin{equation}\label{eq:1}
\Gamma_\text{rad} =\frac{1}{\tau_\text{rad}} =  \frac{n_DE^3_\text{ZPL}\mu^2}{3\pi\epsilon_{0}c^3\hbar^4}\text{,}
\end{equation}
where $\epsilon_0$ is the vacuum permittivity, $\hbar$ is the reduced Planck constant, $c$ is the speed of light, $n_D = 2.1$ is the refractive index of hBN. The calculated radiative lifetime is 11.7~ns which is in the ballpark of PL lifetimes of the observed emitters~\cite{tran2016robust,xu2021creating,schell2017coupling}. With the calculated $S = 2.58$, we simulate the PL spectrum in Fig.~\ref{Figure4}c. The ZPL wavelengths scatter in the range of 430~nm and 490~nm in the observed PL spectra. Although our result could reproduce the first two phonon replicas, the intensity of the simulated phonon sideband is much stronger than that in the experimental spectra. We conclude that the recorded PL spectrum should come from other defects. Further experiments on isolated blue emitters are needed to confirm the existence of \bnnb.

%
%

In conclusion, through first principles calculations, we systematically study the optical properties of native antisite defects in hBN. The calculated ZPL energy and phonon sideband of \bn($0$) agree well with previously reported color centers in hBN. \nb($+$) might be related to recently observed dim blue emitter. Remarkably, we find the out-of-plane distortion exists in native antisite defects. The dynamic instability could dramatically change the optical properties of defects. This leads us to speculate that the out-of-plane distortion might be ubiquitous for many planar defects in hBN. Our result could contribute to the identification and atomic structure analysis of the ongoing exploration of defect emitters in hBN.
\\
\\
\\
\\
\textbf{Competing interests}
The authors declare that they have no competing interests.
\\
\\
\textbf{Data Availability}
Data supporting the findings of this study are available from the corresponding author on a reasonable request.

%
%
\begin{acknowledgments}
A.G. acknowledges the EU HE projects QuMicro (Grant No.\ 101046911) and SPINUS (Grant No.\ 101135699). This research was supported by the Ministry of Culture and Innovation and the National Research, Development and Innovation Office within the Quantum Information National Laboratory of Hungary (Grant No.\ 2022-2.1.1-NL-2022-00004). P.L. acknowledges the NSFC (Grant No.\ 12404094). We acknowledge KIF\"U for awarding us access to high-performance computation resource based in Hungary.
\end{acknowledgments}

\bibliography{mainref}

\end{document}